# Comment on the "Comment by Schwab, Blencowe, Roukes, Cleland, Girvin, Milburn, and Ekinci: quant-phys/0503018"

**A. Gaidarzhy, G. Zolfagharkhani, R.L. Badzey, P. Mohanty**
*Department of Physics, Boston University, Boston, MA 02215*
*Department of Aerospace and Mechanical Engineering, Boston University, Boston, MA 02215*

In a recent publication[1], we reported the observation of discrete displacement of a nanomechanical oscillator in a normal mode frequency of f = 1.5 GHz at a temperature of $T_{low}$ = 110 mK ($kT_{low} \sim hf$), whereas the oscillator displayed continuous displacement at a higher temperature of $T_{high}$= 1000 mK ($kT_{high} >> hf$) according to the Hooke's law, $F \propto kx$. We argue that the oscillator at low temperatures ($k_B T_{low} \sim hf$) displays discrete displacements between two localized states, and the data can be interpreted as transitions between two quantum states.

Schwab *et al.* in their comment[2] on our publication argue against the interpretation of the data that the observed discrete displacement cannot be a signature of a quantum effect. We give a brief response to some of their questions wherever scientifically appropriate. We also assert that the comments are not scientifically justified, and the rhetoric is highly inappropriate.

1) *"According to standard quantum mechanics, measurement of energy quanta ...requires a quantum non-demolition (QND) measurement scheme."*
This statement invalidates 100 years of quantum mechanics experiments, starting from discrete energy spectrum of atoms[3]. Quantum mechanical experiments such as energy quantization and interference effects[a] are routinely done without a QND scheme[3,4,10]. The questionable conjecture of Schwab *et al.* is that the observation of discrete energy spectrum of a quantum system necessarily requires measurement of a specialized non-classical state with minimum uncertainty (via QND). In general, observation of quantum jumps in atomic systems do not involve QND measurement[11,12].
<u>Counter Example</u>: atomic spectra[3], energy spectra of quantum dots or the so-called artificial atoms[5,6,7].

In a separate question, Schwab *et al.* write *"the authors use a room temperature semiconducting amplifier with a noise temperature of $T_N$ = 440 K to detect these voltages [b]. Thus, in addition to the magnetic drive the back action current noise of such an amplifier will drive the resonator, acting as a thermal bath far above temperature of 100 mK quoted by the authors."*
<u>Counter Example:</u> Almost all of mesoscopic physics[9,10] (transport measurements), including quantum hall effect[8] and single-electron transistors[5,7], in which either quantization or interference effects are observed at millikelvin temperatures in presence of preamplifiers with typical voltage noise of 1.1 nV/Hz$^{1/2}$ (equivalent to a noise temperature $T_n$ = 440 K). The argument of Schwab *et al.* suggests that in all these experiments the real temperature due to "the back action current noise" is closer to 440 K rather than in the millikelvin range[9,10]. If the argument of Schwab *et al.* is true, then all of the mesoscopic transport experiments at millikelvin or even kelvin temperatures have to be wrong.

Additional objection by Schwab *et al.*: *"the authors fail to point out that the resonator is driven many orders of magnitude above the ground and first excited state during the measurement…Given the reported parameters ..., the average number of energy quanta in the resonator during the measurement is N = 120,000 >> 1…"*
<u>Counter Example:</u> This comment is erroneous. It suggests that quantum transitions necessarily involve a single quantum of energy (N = 1). Phrased differently, the question is "Could one observe quantum jumps in a quantum system where the jump size between two states involves large numbers (N >> 1) of quanta (such as photons)?" For evidence, we refer to the beautiful experiments by Dehmelt's group[11] and Wineland's group[12] in which the jump size involves thousands of photons (~1000) per second (see, for example, Figs. 2 of Ref. 11 or Ref. 12). The answer is an emphatic yes[c].

---

[a] While energy quantization typically refers to a measurement of the diagonal elements of the reduced density matrix of an open quantum system, interference effects correspond to the off-diagonal elements.
[b] The input noise of our room temperature preamplifier is 1.1 nV/Hz$^{1/2}$, or $T_{noise}$ = 440 K. Even when the impedance mismatches of the transmission line are accounted for, the back action noise incident on the nanobeam is at most 0.3 nV/Hz$^{1/2}$ or equivalently $T_{noise} \sim$ 30 K, clearly still too high.
[c] The argument can also be framed in the context of "energy quantization in a quantum dot" in a so-called artificial atom or single-electron transistor. The size of the source-drain current measured in a quantization experiment typically involves a large number of electrons. Such measurements allow observation of usually a billion electrons through the quantum dot (island) at a time. Only recently there have been time-resolved measurements of transport through quantum dots[14].

2) "*For Q ~ 100 and ω = $10^{10}$ $s^{-1}$, the average lifetime of an energy quantum is ~ 10 nsec. Even if the authors could measure the oscillator energy with single quantum accuracy, the observed jumps due to decay would certainly not be as long as tens of minutes…a discrepancy of over 10 orders of magnitude from expectation*".

The alleged lifetime of 10 ns is extracted from $\tau = Q/\omega$, which is the classical ring-down time of a classical oscillator with frequency $\omega/2\pi$ and quality factor Q. This is not the implied lifetime $T_1$ of a quantum system with discrete energy levels. Furthermore, $T_1$ of different energy levels in an atom is different for different energy levels[3], spanning a broad range of timescales, from tens of nanoseconds to seconds or minutes (metastable states). In a transition with multiple quanta the decay time observed in a statistical measurement will depend on the distribution of $T_1$ for all energy levels. Fundamentally, the decay of an atom could be observed in any timescale, since the exponential decay form extends all the way to infinity. Only the number of counts at longer timescales becomes exponentially small.

Finally, in the paper we never mentioned anything with regard to single quantum accuracy.

3) "*The magnetomotive response of the suspected 1.48 GHz mode is anomalous. Fig#3b shows the amplitude versus magnetic field where the authors claim that it demonstrates the expected quadratic dependence. The authors fail to point out and explain that the quadratic fit is not symmetric about the origin as expected and widely observed...*"

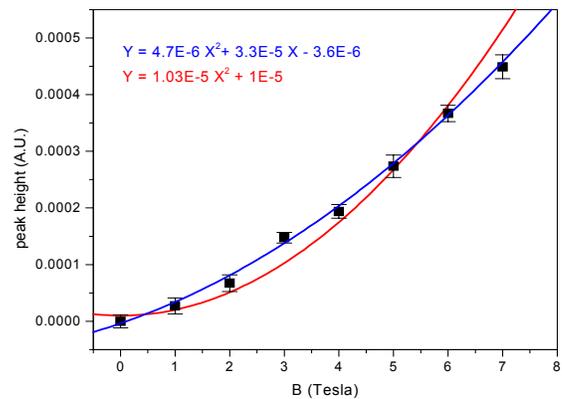

The fit to the data in Fig. 3b is indeed a quadratic fit (quadratic defined as $y = a_0 + a_1 B + a_2 B^2$), as we clearly mention in the caption. This is different from a fit to $B^2$. The difference is the asymmetric linear term, applied to the data on the positive magnetic field. We reproduce here the original data with both fits. The unbiased quadratic fit shown in the original plot is the blue curve, which is shifted from the origin by 2 Tesla. The red curve is a fit to $y = a_0 + a_2 B^2$, symmetric about the origin B = 0. Both curves are within the measurement noise of our data, and the peak response height grows as $B^2$, in accordance with the magnetomotive technique. The discrepancy in the coefficient of the quadratic term is of no consequence to the rest of the paper, because we derive values for the stiffness of the mode directly from force vs. displacement measurements.

4) "*Finally, there is no justification offered for the assertion that the motion of the central beam is amplified from femtometers to picometers in comparison to the femtometer motion of the fingers at 1.5 GHz.*"

The detected signal, $V_{emf}$, is indeed amplified by 2-3 orders of magnitude in comparison to an equivalent signal $V_{emf}$ from a 1.5-GHz straight beam with femtometer displacements. There are three distinct contributions to the signal amplification.
(i) We specifically designed the multi-element structure to have a low effective spring constant $k_{coll}$ in the collective mode. In our oscillator, $k_{coll}$ is in fact measured to be lower than the spring constant of a 1.5-GHz straight beam by an order of magnitude[1]. This results in an enhanced displacement by an order of magnitude.
(ii) Induced voltage $V_{emf}$ in the detection electrode depends on the enclosed flux, $V_{emf}$ = Re (dΦ/dt) ~ ηBL (dy/dt). The area swept by the central beam is roughly equal to the product of length (L) and the displacement of the central part of the beam. The longer the beam, the larger the area or the induced emf. Compare it to an equivalent 1.5-GHz straight beam with a length of 1 micron. The deflection of the central part of the 1.5-GHz straight beam (amplitude) will be less because of high stiffness constant k (part (i)). Furthermore, the length L is 10 times shorter. The net enhancement in the swept area is roughly a factor of 100.
(iii) The striking aspect of the antenna structure is that it is not a single cantilever. It consists of 40 single cantilevers, 20 on each side. As we describe in the paper, all cantilevers move coherently in phase in the collective mode. A system of coherently coupled oscillators, quantum or classical, in certain symmetry, can be described by a collective coordinate with an effective displacement greater than the displacement of an individual oscillator[13]. The motion of the central beam can be described in a basis of collective coordinates that demonstrates the enhancement of displacement. We will have a detailed description of our novel structure design in a future publication.

In summary, both the objections and the premise of the comment, on the data interpretation, by Schwab *et al.* are not valid. As we clearly mention in the paper, a proper theoretical framework to understand quantized motion of a macroscopic mechanical system of 50 billion atoms, in presence of decoherence and dissipation, is yet to be developed.

The observed phenomenon is robust and reproducible. We will be glad to provide samples of antenna structures to any group for repeating this experiment. We will also host researchers to carry out this experiment in our laboratory, if they do not have access to high frequency measurement and low millikelvin temperature setups[d].

---

[d] Email: mohanty@physics.bu.edu for details.